\documentclass[10pt,letterpaper,twocolumn]{article} 

\usepackage{ol2}
\usepackage[draft]{hyperref}
\usepackage{amsmath}

\begin{document}

\twocolumn[ 

\title{Tunable frequency-stabilization of UV laser using a Hallow-Cathode Lamp of atomic thallium}


\author{Tzu-Ling Chen$^1$, Chang-Yi Lin$^1$, Jow-Tsong Shy$^{1,2}$ and Yi-Wei Liu$^{1,*}$}
\address{$^1$Department of Physics, \\ National Tsing Hua University, Hsinchu, Taiwan 30013}
\address{$^2$Institute of Photonics Technologies, \\ National Tsing Hua University, Hsinchu, Taiwan 30013}
\address{$^*$Corresponding author: ywliu@phys.nthu.edu.tw}

\begin{abstract}A frequency-stabilized ultraviolet laser system, locked to the thallium resonant transition of 377.5~nm, was demonstrated using a novel bichromatic spectroscopy technique for tuning the zero-crossing laser-lock point. The atomic thallium system is a promising candidate in atomic parity violation and permanent electric dipole moment experiments, and its 377.5~nm ${6P_{1/2}\rightarrow7S_{1/2}}$ transition is important for thallium laser cooling and trapping experiment. The pressure shift, owing to the high pressure buffer gas of  the hollow-cathode lamp, was observed using an atomic beam resonance as reference. Such a shift was corrected by adjusting the peak ratio of the two Doppler-free saturation profiles resulted from two pumping beams with a 130~MHz frequency difference.  The resulted frequency stability of the ultraviolet laser is $\sim$0.5~MHz at 0.1~sec integration time. This scheme is compact and versatile for stabilizing UV laser systems, which acquire a sub-MHz stability and frequency tunability.

\end{abstract}


 ] 

\maketitle 

Laser stabilization on atomic or molecular reference transitions is essential to high precision spectroscopy and laser cooling experiment. For the experiments with low vapor pressure samples, a gas cell as a reference requires a high temperature oven or a beam machine.  For example, atoms with high melting points \cite{Ross:1995wk}, such as atomic strontium \cite{PhysRevLett.91.243002}, ytterbium \cite{Bowers:1999fg} and thallium (Tl) \cite{Chen:2012iu}, would require a 300-900$^\circ \rm C$  cell temperature to reach a sufficient number density. Hollow-cathode lamp (HCL), which provide high number densities of atoms or molecules in the ground or excited states using electrical discharge,  is a versatile and compact setup for such low vapor pressure spices.  Various spectroscopy techniques have been implemented using HCLs to stabilize laser frequency with atomic ytterbium \cite{Kim:2003vm}, calcium \cite{Dammalapati:2009bt}, and thorium \cite{DeGraffenreid:2012wv}. 
However, owing to the high buffer gas pressure in the HCL, a shift and broadening of the spectra \cite{Masaki:1988wm}, which are important as a frequency reference, need to be carefully studied and compensated using a frequency correction in many applications.

In our experiment, a 377.5~nm frequency-doubled Ti:Sapphire laser was locked to the Doppler-free saturation spectroscopy of atomic Tl in a HCL using the bichromatic spectroscopy technique \cite{Van:2004v, Chou:2004jn}. To study the pressure shift, an atomic Tl beam system was simultaneously used to observe the fluorescence spectrum as the shift-free reference for comparison with HCL. A simple method of correcting the frequency shift to the transition center was also demonstrated. 

Atomic Tl is an attractive element for atomic parity violation (APV)  and permanent electric dipole moment (EDM) experiment mainly attributing to its large atom number (Z=81), which enlarge APV and EDM effects by $\rm{Z^3}$ \cite{Ginges:2004wa}. Its strong ${6P_{1/2}\rightarrow7S_{1/2}}$ resonant transition, at 377.5~nm, permits optical pumping to the metastable state ${6P_{3/2}}$ prepared for the nearly closed transition cycle ${6P_{3/2} (F=2) \leftrightarrow 6D_{5/2} (F=3)}$ for the laser cooling and trapping experiments \cite{Fan:2011bf}, which can greatly reduce the systematic shifts and the spectral-broadening effects in potential applications of high-resolution spectroscopy. Due to the high melting point (304$^\circ\rm C$), atomic Tl HCL is versatile to stabilize the 377.5~nm UV laser for the Tl laser cooling experiment.

\begin{figure}[hbt]
\centerline{\includegraphics[width=1\linewidth]{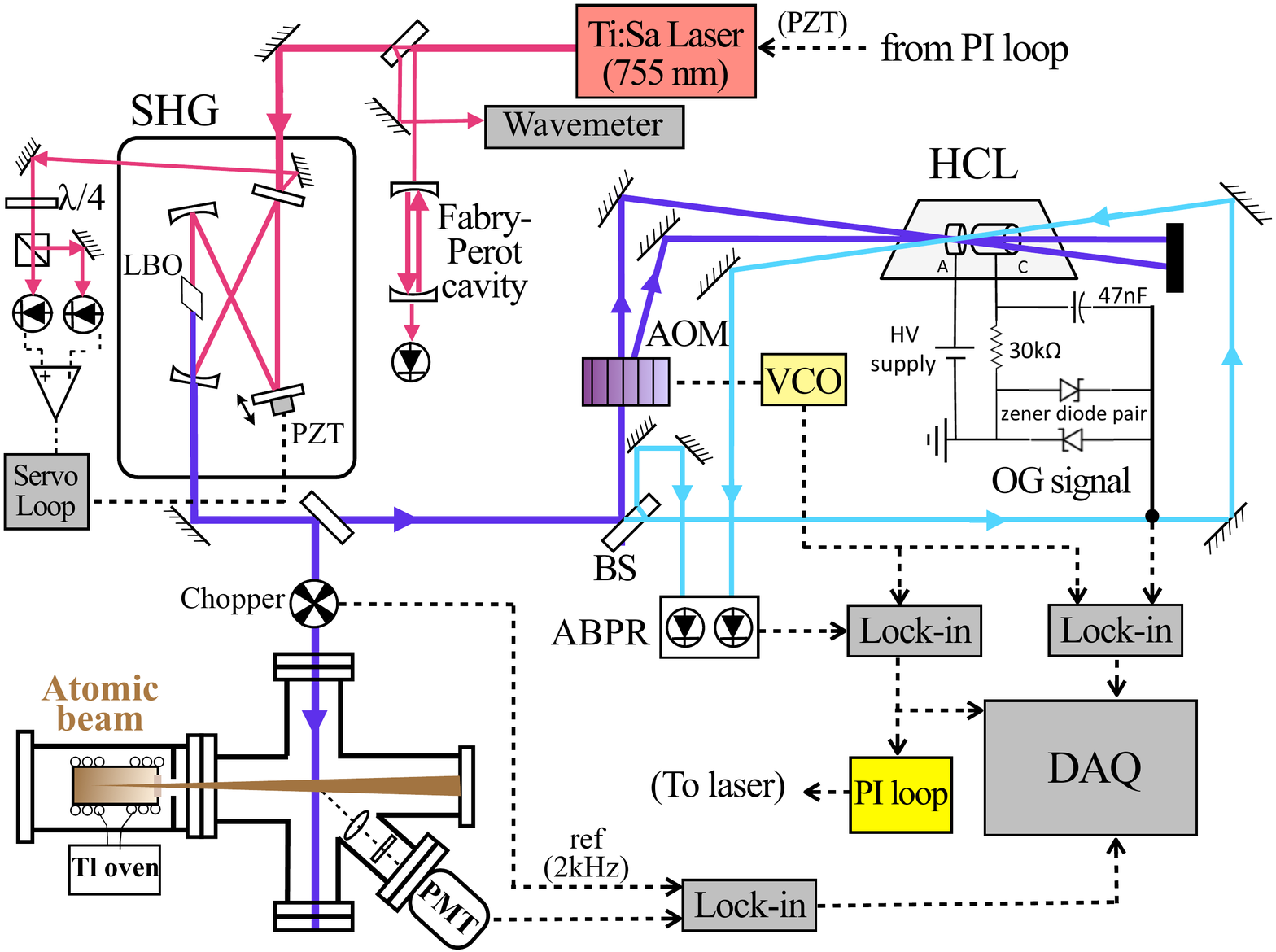}}
\caption{\label{fig:setup}(Color Online) Experimental setup.
BS, beam splitters; ABPR, auto-balanced photo receiver. SHG: second harmonic generation. AOM: acousto-optic modulator. DAQ: data acquisition. VCO: voltage-controlled oscillator. PI: proportional-integral}
\end{figure}

The experimental apparatus is shown schematically in Fig.~\ref{fig:setup}. It includes three main parts : HCL spectrometer, atomic beam spectrometer and 377.5~nm laser system. The HCL (Hamamatsu L2783-81NE-TL) is a see-through type with a ring cathode length of 19~mm and a bore diameter of 3 mm. It  is filled with 14~Torr neon as buffer gas. The nominal discharge current is 9~mA, corresponding to 490~V voltage drop across the discharge with a 30~k $\Omega$ ballast resistor.
The 377.5~nm laser system includes a single frequency cw Ti:sapphire laser operated at 755~nm and a frequency-doubler based on a traveling-wave ring cavity with a Brewster-angle cutted LBO crystal.
Total power of the generated UV light is typically 1~mW. One part ($\sim$300~$\mu$W) of the UV light was sent to the Tl atomic beam system for the sub-Doppler fluorescence spectrum as a shift-free frequency reference at  ${6P_{1/2}\rightarrow7S_{1/2}}$. Another part ($\sim$500~$\mu$W) was to the bichromatic spectrometer for the frequency stabilization.

For the atomic beam spectroscopy, the UV light was modulated at 2~kHz by a mechanical chopper and interacted perpendicularly with an aperture-collimated atomic beam with a 90~mrad divergent angle. The atomic beam apparatus include a stainless chamber with UV AR coated windows and with background pressure of 10$^{-5}$ Torr. The atomic beam was generated by heating a bulk of Tl to 420$^{\circ}$C. The laser induced fluorescence was detected by a photomultiplier and demodulated using a lock-in amplifier. The details of the experiment can be found in our previous report \cite{Chen:2012iu}. A temperature-controlled Fabry-Perot interferometer with a free spectral range of 300~MHz (in 755~nm) was used to diagnose the laser scanning and calibrate the scanned frequency range.

In order to provide a Doppler-free dispersion-like signal for laser stabilization, the bichromatic spectroscopy and the Doppler-free modulation transfer technique are implemented in the HCL spectrometer. In Fig.~\ref{fig:setup}, the UV laser was separated into two pumping beams and a counter-propagating probe beam. The dichromatic spectroscopy technique requires two pump beams with a frequency difference, which was provided using an AOM (Intraaction) driven by a 260~MHz radio-frequency with a 90~kHz amplitude-modulation. 
The 0th ($\sim$180~$\mu$W) and the 1st ($\sim$175~$\mu$W) order diffractions from AOM, as pump beams, generated two saturated absorption profiles, which were 260/2=130~MHz separated. The differential signal of the two saturation profiles, given by the output of the lock-in amplifier, was a dispersion-like profile to serve as an error signal for laser stabilization. The error signal passed through a Proportional-Integral (PI) filter to the piezo of the Ti:Sapphire laser cavity.
For the power fluctuation normalization, a portion of UV laser beam ($\sim$100~$\mu$W) was sent to an auto-balanced photo receiver (ABPR, Nirvana Model 2007) , which is capable of reducing common mode noise by over 50dB at the noise spectrum of DC to 125 kHz. The output signal of ABPR was then demodulated using a lock-in amplifier.  Meanwhile the ac signal of the discharge current, the optogalvanic effect, induced by the laser intensity modulation was detected by a 30~k $\Omega$ ballast resistor through a 47~nF coupling capacitor.

\begin{figure}[hbt]
\centerline{\includegraphics[width=1\linewidth]{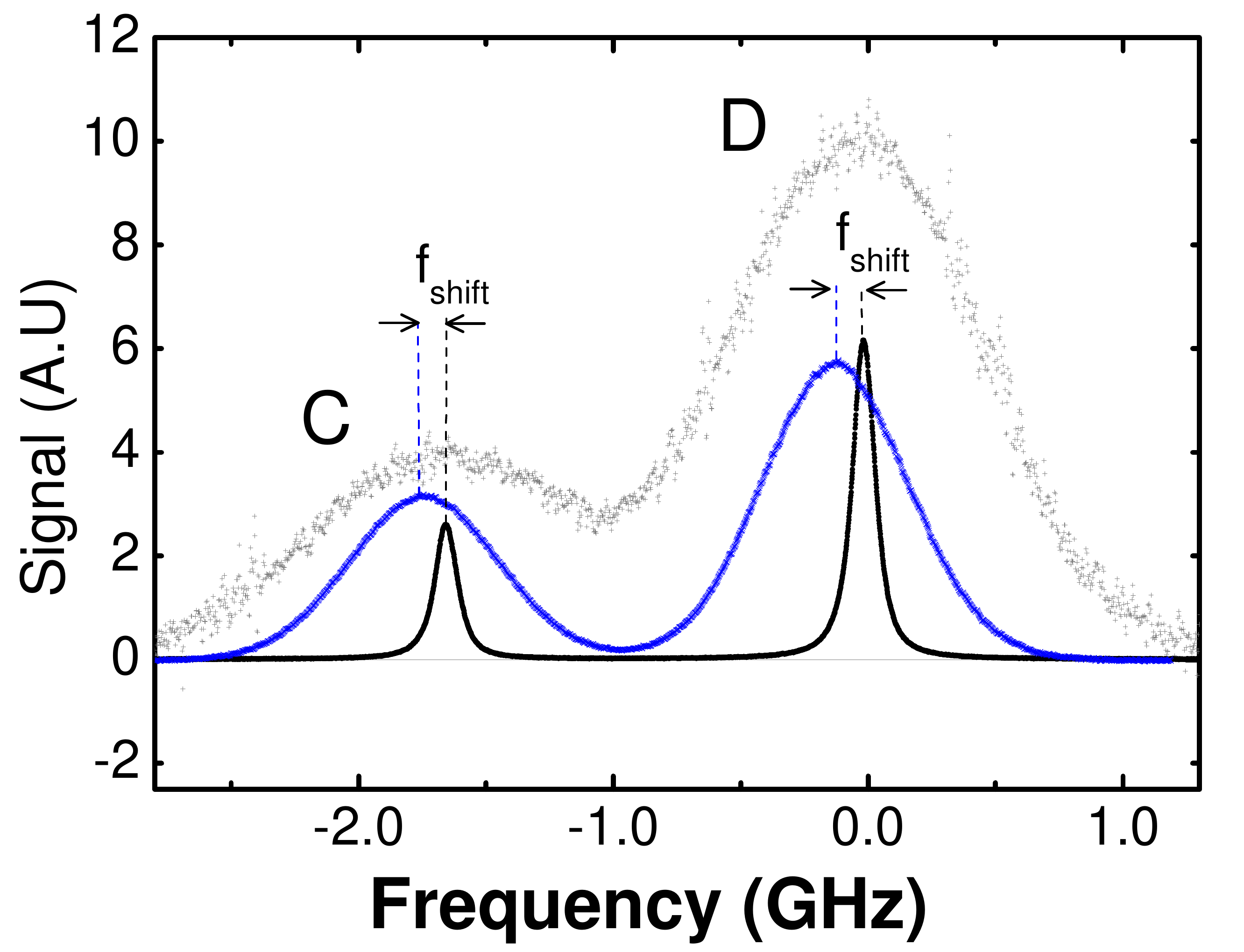}}
\caption{\label{fig:SA}(Color Online)Typical spectra of thallium ${6P_{1/2},F=1\rightarrow7S_{1/2},F=1}$ transition of ${^{203}}$Tl (C) and ${^{205}}$Tl (D). 
Gray dots: optogalvanic spectrum in HCL. Blue dots: saturation absorption spectrum in HCL. Black dots: Doppler shift-free laser induced fluorescence spectrum in atomic beam system. There is a frequency shift between the spectra in HCL and atomic system.}
\end{figure}

Figure~\ref{fig:SA} shows the hyperfine transition ${6P_{1/2},F=1\rightarrow7S_{1/2},F=1}$ at 377.5~nm, where the HCL Doppler-free saturation absorption spectrum (blue dots), the Doppler broadened optogalvanic spectrum (gray dots) and the atomic beam fluorescence spectrum (black dots) were recorded simultaneously. The HCL saturation absorption signal was obtained by the modulation transfer technique with only the 1st order pump beam. 
The atomic beam fluorescence spectrum was fitted by a Voigt function with a residual Doppler width $\sim$ 70~MHz. In comparison with the atomic fluorescence spectrum, a relative frequency shift of  both the HCL saturated absorption peak and optogalvanic peak was found. 
This 100~MHz red shift (f$_{\bf{shift}}$ shown in the Fig.~\ref{fig:SA}), which occurred on both two isotopes and all the hyperfine transitions, was contributed mainly from the pressure shift induced by Ne buffer gas\cite{Dygdaa:1999ws} in HCL. A part of the shift could be contributed to the first order Doppler shift of the atomic beam spectrometer, because of the non-perpendicularity between the atomic beam and the laser beam.

\begin{figure}[hbt]
\centerline{\includegraphics[width=1\linewidth]{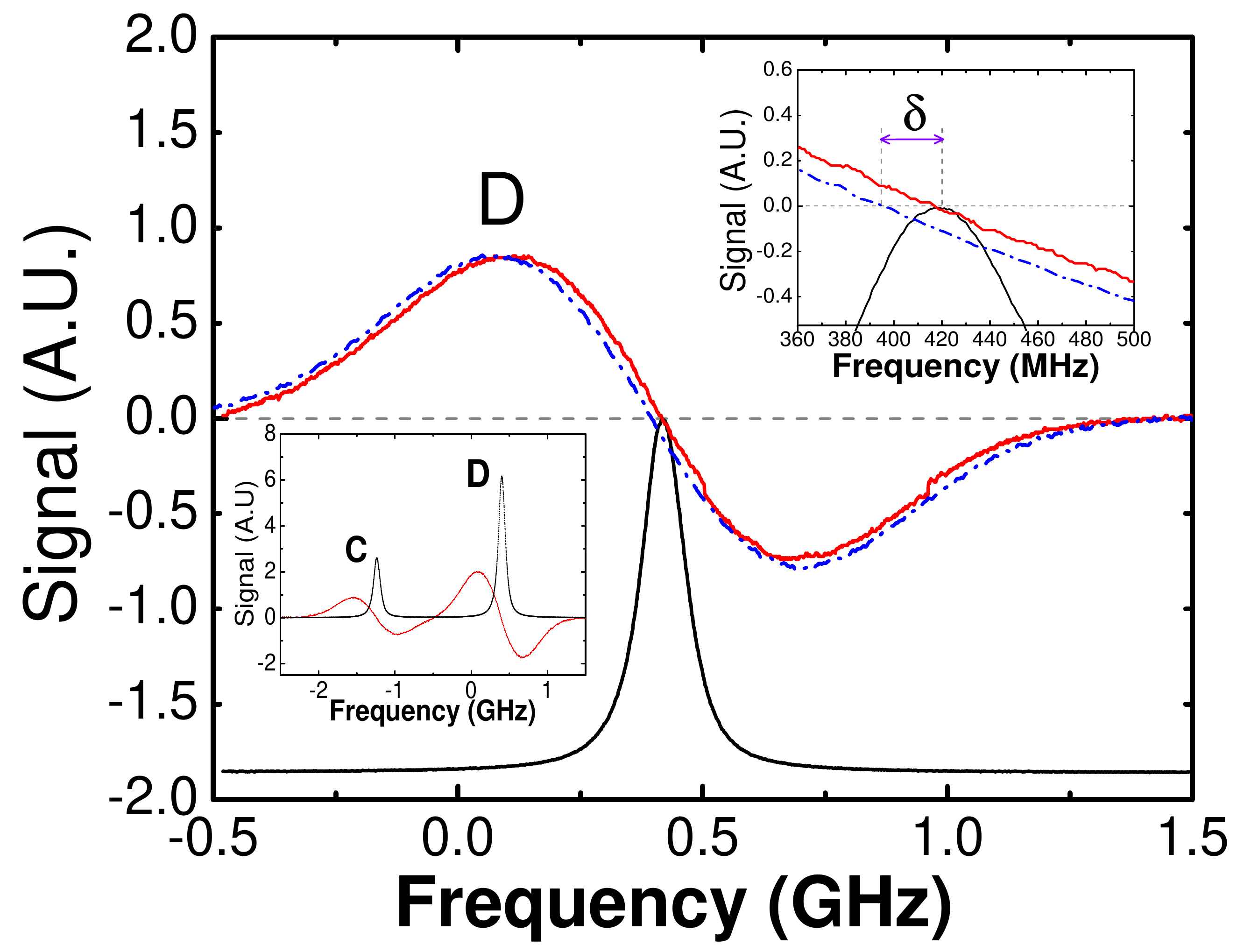}}
\caption{\label{fig:dispersion}(Color Online)The tunable laser-lock point of dispersion-like signal in bichromatic spectroscopy.
The blue dashed-dotted curve shows the original shape before adjustment and the red solid curve shows the shifted zero-crossing point after adjustment. Upper inset: zoom in the zero-crossing point. Lower inset: the overall bichromatic spectra of $^{203}$Tl (C) and $^{205}$Tl (D).}
\end{figure}

The lock-in amplifier output is the differential signal between the in- and out-phase signals, referenced to the modulation. In this experiment, the resulted signal was the difference of the two saturation dips from the 1st and 0th order pump beams. A simple model of such dispersion-like signal can be written as :
$$
S(\omega)=\rm{A_0}G(\omega-\omega_0)-\rm{A_1}G(\omega-(\omega_0-\Delta/2)),
$$

where $\rm{A_0}$ and $\rm{A_1}$ are the amplitude-modulation coefficients of the pump beams of the 0th and 1st order, respectively. The parameter $\Delta$ is the rf frequency of the AOM. Because of a high buffer gas pressure in the HCL, the observed single-beam lineshape under the effect of velocity-changing collisions \cite{Richardson:1987wm} is gaussian G$(\omega-\omega_0)=$exp$(-(\omega-\omega_0)^2/2\sigma^2)$. Hence, the zero-crossing point S($\omega_{zero}$)=0, where the laser locked, is at the frequency:
$$
\omega_{zero}=\omega_0-\frac{\Delta}{4}+\frac{2\sigma^2}{\Delta}{\bf{ln}}(\frac{A_0}{A_1})
$$
In the case of perfect overlapping of the three beams (two pump beams and one probe beam),  $\rm{A_0}=\rm{A_1}$ and the zero-crossing point is at $\omega_{zero}=\omega_0-\Delta/4$. While one of the pump beams is slightly mis-aligned or attenuated, a non-equal amplitude modulation $\rm{A_0}\neq\rm{A_1}$ results an asymmetrical dispersion-like profile, which has a shifted zero-crossing point.  In our experiment, this feature was utilized to tune the frequency of the locked laser by varying the overlapping of the laser beam and compensate the pressure shift in the HCL to set the zero-crossing point on the resonance frequency of the atomic beam. In the Fig.~\ref{fig:dispersion}, before adjustment (blue dashed-dotted curve) there is a offset ${\delta}$$\sim$30~MHz between the zero-crossing point and fluorescence peak. By adjusting the alignment of one of the pump beam (the red solid curve in the inset of Fig.~\ref{fig:dispersion}), the zero-crossing point was shifted to match the fluorescence peak.
The peak-to-peak width of the line was measured to be 1.2 GHz. The slope of the first derivative curve used as the laser-lock signal was 0.52~mV/MHz at the center frequency and the signal to noise ratio was estimated to be 350 at an integration time of 0.1~s.

\begin{figure}[hbt]
\centerline{\includegraphics[width=1\linewidth]{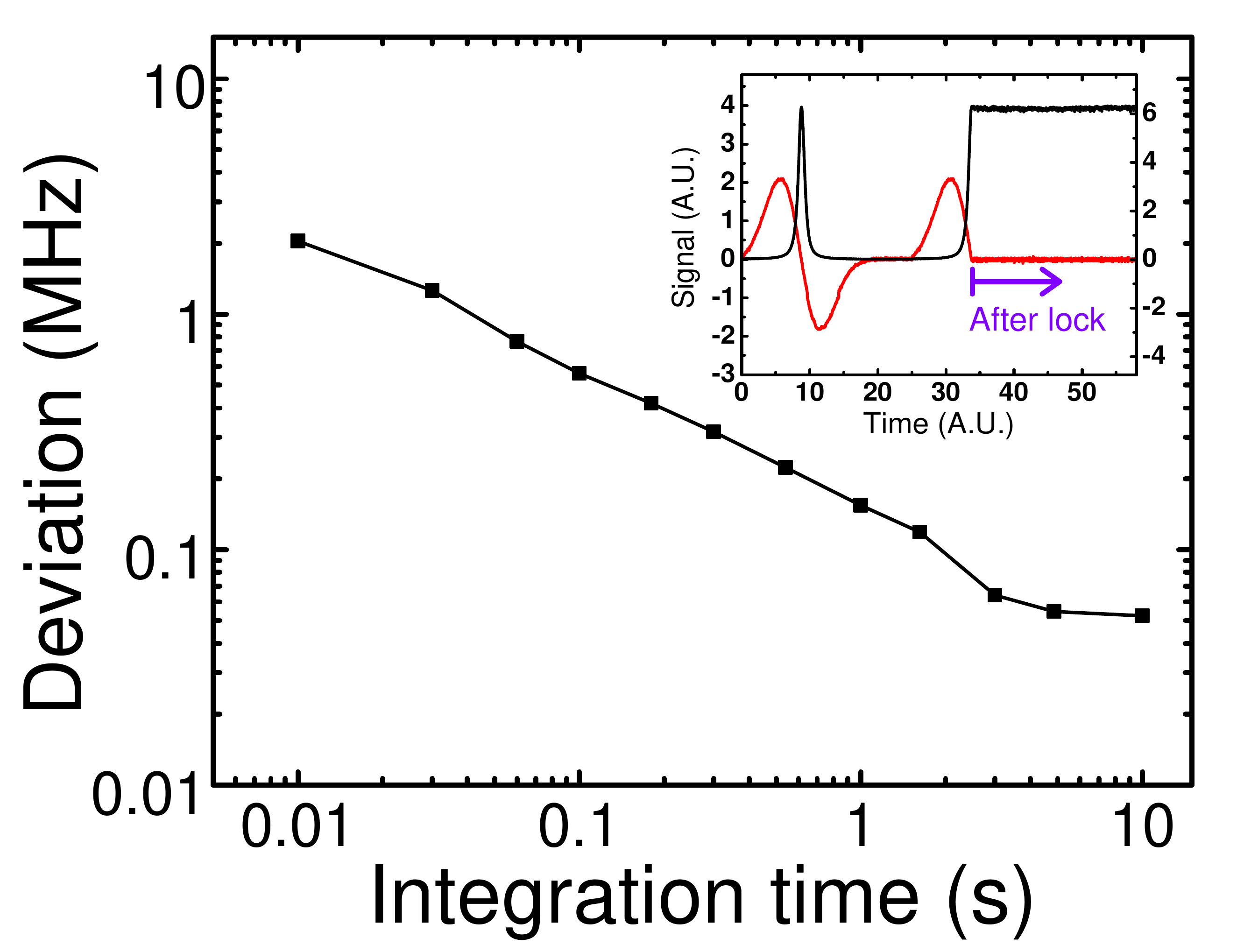}}
\caption{\label{fig:Allen}(Color Online)Allen deviation. Inset: Recording of the error signal (red) and fluorescence signal fluctuation (black) with the locked laser.}
\end{figure}

Figure~\ref{fig:Allen} shows the Allen deviation for the frequency-stabilized Ti:Sa at various integration times. The error signal of the locked laser shown in the inset was used to evaluate the stability of the laser frequency. The frequency stability was 500~kHz at 0.1~s integration time and reached a value of 50~kHz at 10~s. In the inset of Fig.~\ref{fig:Allen}, it shows that the laser was locked onto the fluorescence peak to maintain a continue excitation of the ${6P_{1/2}\rightarrow7S_{1/2}}$ transition. This stability can be maintained for a time $>$ 500~sec in typical experimental condition.

In conclusion, we have demonstrated a tunable zero-crossing laser-lock point of dispersion-like signal using bichromatic spectroscopy within a compact Tl HCL. The demonstrated method, in which the zero-crossing point is shifted to compensate the relative frequency shift of the saturation profile and the resonance line, provides a precise locking to an atomic resonance of interest.
The frequency stability with a deviation of 0.5~MHz at a 0.1~sec integration time was achieved, which is sufficiently narrow to be used as a compact frequency reference for the future laser cooling experiments of various atomic species, including atomic thallium.

The authors thank Nang-Chian Shie for his useful advice and help in the initial construction of the HCL system. This research was supported by the National Science Council of 410 Taiwan under Grants No.100-2112-M-007-006-MY3 and No. 411 99-2112-M-007-001-MY3.

\bibliographystyle{ol}
\bibliography{OGE_ref}

\end{document}